\documentclass[aps,prl,twocolumn,groupedaddress]
{revtex4}
\usepackage{graphicx} 
\usepackage{dcolumn} 
\usepackage{longtable} 
\usepackage{amssymb} 
\usepackage{amsmath} 
\usepackage{amsfonts} 
\usepackage{slashed} 
\usepackage[dvipsnames]{xcolor} 
\def\lamb#1#2{$^{#1}_{\Lambda}${#2}}

\def\lamblamb#1#2{$^{~~#1}_{\Lambda\Lambda}${#2}}

\newcommand{\bs}[1]{\boldsymbol{#1}} 
\newcommand{\nopieft}{\mbox{$\slashed{\pi}$EFT}}

\newcommand{\be}{\begin{equation}} 
\newcommand{\ee}{\end{equation}} 
\newcommand{\rvec}{{\bs{r}}}

\newcommand{\half}{\frac{1}{2}} 

\begin{document} 
   
\title{In-medium $\Lambda$ isospin impurity from charge symmetry breaking in 
the \lamb{4}{H}--\lamb{4}{He} mirror hypernuclei} 
\author{M.~Sch\"{a}fer} \author{N.~Barnea} \author{A.~Gal}
\affiliation{Racah Institute of Physics, The Hebrew University, 
Jerusalem 91904, Israel} 

\date{\today}

\begin{abstract} 
The $\Lambda$ separation energies in the mirror hypernuclei 
\lamb{4}{H}--\lamb{4}{He} exhibit large charge symmetry breaking (CSB). 
Analyzing this CSB within pionless effective field theory while using 
partially conserved baryon-baryon SU(3) flavor symmetry, we deduce 
a $\Lambda -\Sigma^0$ induced in-medium admixture amplitude ${\cal A}_{I=1}
\approx 1.5\%$ in the dominantly isospin $I=0$ $\Lambda$ hyperon. 
Our results confirm the free-space value ${\cal A}^{(0)}_{I=1}$ inferred 
directly within the SU(3) baryon octet by Dalitz and von-Hippel in 1964 and 
reaffirmed in a recent QCD+QED lattice calculation. Furthermore, exploring the 
consequences of SU(3) flavor symmetry on the $\Lambda$-nucleon interaction, 
we find that CSB is expected to impact the $S=1$ and $S=0$ spin channels in 
opposite directions, with the latter dominating by an order of magnitude. 
These observations explain a recent deduction of $\Lambda$-nucleon CSB 
strengths. 
\end{abstract}

\maketitle

\noindent 
{\bf Introduction.}~
Renormalization of hadron decay constants in nuclear matter is a recurring 
theme in hadronic physics. Well known examples are the roughly 30\% in-medium 
quenching of the pion decay constant $f_{\pi}$ and the weak-decay axial-vector 
constant $g_A$. The quenching of $f_{\pi}$ was inferred from deeply bound 
$\pi^-$-atom levels in heavy nuclei~\cite{KY01}, soon shown by Friedman to 
hold over the whole periodic table~\cite{F02}, in line with a chiral-symmetry 
partial restoration argument by Weise~\cite{Weise00}. The quenching of $g_A$ 
was noticed by Wilkinson in nuclear Gamow-Teller $\beta$ decays~\cite{Wilk73} 
and soon attributed by Rho~\cite{Rho74} to spin-isospin correlations manifest 
in the Ericson-Ericson-Lorentz-Lorenz renormalization of the $p$-wave 
pion-nucleus optical potential \cite{EE66}.  

Less explored is the ${\cal S}=-1$ strange hadronic sector, where the poorer 
data base of $\Lambda$ hypernuclei~\cite{GHM16} limits the deduction of 
in-medium trends. A particularly interesting question is whether and how 
far the Dalitz-von Hippel (DvH)~\cite{DvH64} relatively large amplitude  
${\cal A}^{(0)}_{I=1}\approx 1.5\%$ of an $I$=1 admixture in the dominantly 
$I$=0 $\Lambda$ hyperon gets renormalized in dense matter. 
${\cal A}^{(0)}_{I=1}$ was inferred by DvH from the $\Lambda -\Sigma^0$ 
mass-mixing matrix element $M_{\Sigma^0\Lambda}$ related in SU(3)-flavor 
(SU(3)$_{\rm f}$) symmetry to octet baryon electro-magnetic mass differences 
$\delta M_{BB'}=M_B-M_{B'}$: 
\begin{equation}   \label{eq:DvH1} 
  M_{\Sigma^0\Lambda}=\frac{1}{\sqrt 3}  
  (\delta M_{\Sigma^0\Sigma^+}-\delta M_{np})=1.14\pm 0.05~{\rm MeV}, 
\end{equation} 
leading to the free-space value 
\begin{equation} 
{\cal A}^{(0)}_{I=1}=M_{\Sigma^0\Lambda}/\delta M_{\Lambda\Sigma^0}=-0.0148
\pm 0.0006. 
\label{eq:DvH2} 
\end{equation} 
We note that this result holds also in quark models, where SU(3) octet 
and decuplet baryons are assigned to the {\bf 56} SU(6) multiplet, by assuming 
only one-quark mass terms and two-quark interaction terms \cite{GS67} but 
no three-quark interaction terms. More recently, Eq.~(\ref{eq:DvH2}) was 
confirmed in a QCD+QED lattice calculation~\cite{LQCD20}, although with 
considerably larger uncertainty of order 30\%. 

\begin{figure}[!t] 
\begin{center} 
\includegraphics[width=0.48\textwidth]{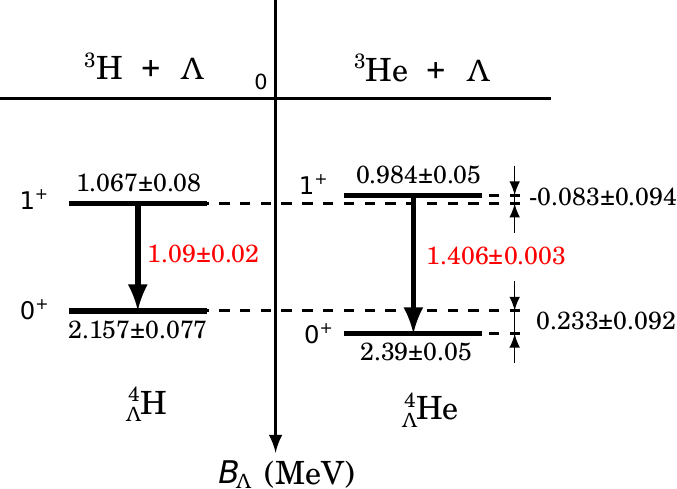} 
\caption{$A=4$ hypernuclear level scheme \cite{E13,MAMI16a,MAMI16b} 
with $\gamma$-ray energies \cite{E13} marked in red. CSB splittings 
are shown in MeV to the right of the \lamb{4}{He} levels. 
Figure adapted from Ref.~\cite{MAMI16b}.} 
\label{fig:A=4} 
\end{center} 
\end{figure} 

Appreciable $\Lambda N$ charge symmetry breaking (CSB) is implied by this 
isospin impurity of the $\Lambda$ hyperon. Unfortunately, the poorly known 
two-body $\Lambda N$ scattering data are limited to $\Lambda p$. 
In $\Lambda$ hypernuclei~\cite{GHM16}, CSB affects mirror states 
($N\leftrightarrow Z$), e.g., the 0$^+_{\rm g.s.}$ and 1$^+_{\rm exc.}$ 
\lamb{4}{H}-\lamb{4}{He} levels where differences of $\Lambda$ separation 
energies $B_{\Lambda}(J^\pi)$ are nonzero: $\Delta B_{\Lambda}(0^+_{\rm g.s.})
$=233$\pm$92~keV and $\Delta B_{\Lambda}(1^+_{\rm exc.})$=$-$83$\pm$94~keV, 
see Fig.~\ref{fig:A=4}. A particularly precise measure of CSB is given by 
their difference, $\Delta \Delta B_{\Lambda}$=316$\pm$20~keV, equal to the 
difference $\Delta E_{\gamma}$ between the two $\gamma$ ray energies marked 
red in the figure. This value is about four times larger than the nuclear CSB 
splitting $\Delta B_{\rm CSB}(^3$H-$^3$He)=67$\pm$9~keV in the mirror core 
nuclei $^3$H-$^3$He \cite{MM01}. The $A$=4 hypernuclear CSB $B_{\Lambda}$ 
splittings have been studied recently within chiral effective field theory 
($\chi$EFT) at leading-order (LO) \cite{GG16} and next-to-leading-order (NLO) 
\cite{HMN21}. In the latter work, the \lamb{4}{H}-\lamb{4}{He} splittings 
were used to estimate the impact of CSB on the $\Lambda N$ interaction.
It was found that CSB acts predominantly on the spin $S$=0 channel and that 
it affects the $S$=1 and $S$=0 channels in opposite directions, but no 
satisfactory theoretical argument was given to motivate this difference. 
Below, we show that these findings are consequences of SU(3)$_{\rm f}$.

\begin{figure}[!t] 
\begin{center}  
\includegraphics[width=0.48\textwidth]{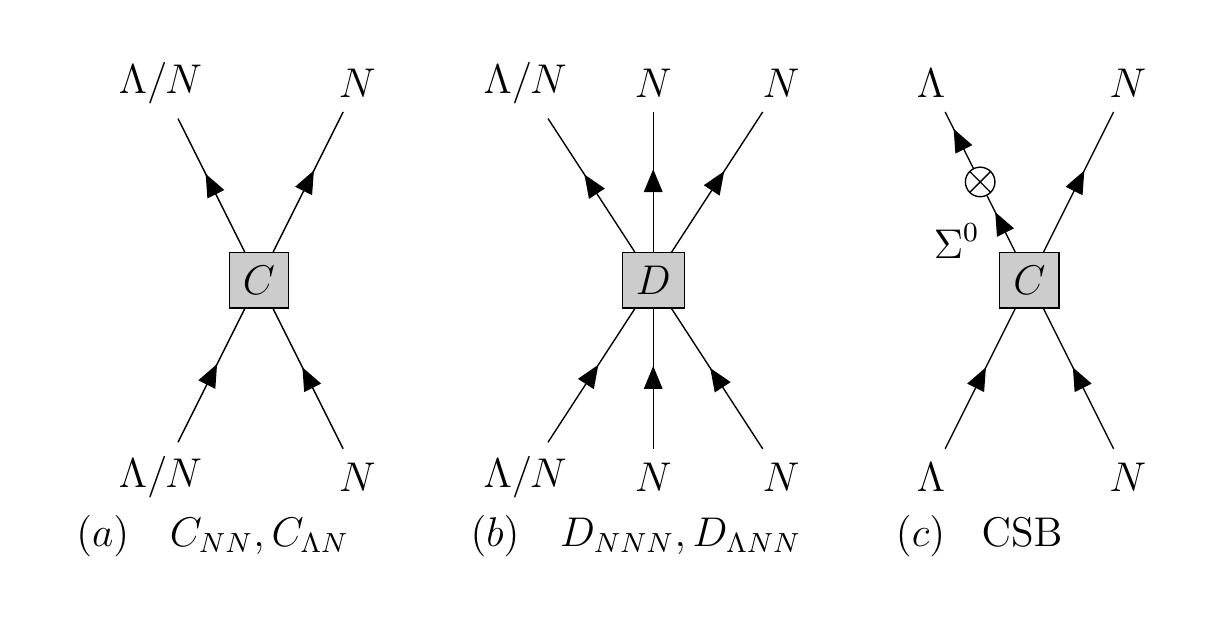} 
\caption{$\Lambda$ hypernuclear CS two-body (a) and three-body (b) 
contact-term diagrams and their associated low-energy constants $C$ and $D$, 
respectively, plus in (c) a $\Lambda N\to\Sigma^0 N$ contact-term diagram, 
followed by a cross for $\Lambda-\Sigma^0$ mixing, to illustrate 
a \nopieft(LO) realization of the CSB ansatz (\ref{eq:DvH3}).} 
\label{fig:LECs} 
\end{center} 
\end{figure}

It was noted by DvH that $\Lambda-\Sigma^0$ mixing induces a long range, 
one-pion exchange (OPE), CSB $\Lambda N$ potential. This result was 
generalized in Ref.~\cite{gal15} relating the CSB $\Lambda N$ 
potential to the charge-symmetric (CS) $\Lambda N\leftrightarrow\Sigma N$ 
transition potential,
\begin{equation} \label{eq:DvH3} 
  \langle \Lambda N|V_{\rm CSB}|\Lambda N\rangle = 
  -\frac{2}{\sqrt{3}}\,{\cal A}^{(0)}_{I=1}\,
       \langle\Sigma N|V_{\rm CS}|\Lambda N\rangle\,\tau_{Nz}.  
\end{equation} 
A schematic illustration of this CSB ansatz is given in Fig.~\ref{fig:LECs}, 
diagram (c). The factor 2 emerges from applying $\Lambda-\Sigma^0$ mixing
to either incoming or outgoing $\Lambda$ states. Projecting the $I_{YN}=1/2$ 
$\Sigma N$ state on the r.h.s. onto its $\Sigma^0 N$ component produces the 
factor $-\tau_{Nz}/\sqrt{3}$, with $\tau_{Nz}=\pm 1$ for $p,n$ respectively. 

Here we explore whether the CSB ansatz (\ref{eq:DvH3}) is satisfied in the 
\lamb{4}{H}-\lamb{4}{He} mirror hypernuclei, i.e., to what extent the value 
of ${\cal A}^{(0)}_{I=1}$ is renormalized in matter. To this end we introduce 
CSB into pionless effective field theory (\nopieft) studies of few-body 
hypernuclei at LO \cite{cbg18,csbgm19}. Since $\Lambda N$ one-pion 
exchange (OPE) is forbidden for the dominantly $I$=0 $\Lambda$ hyperon, 
a \nopieft~breakup scale $2m_{\pi}$ is assumed, remarkably close to the 
threshold value $p_{\Lambda N}^{\rm th}\approx 283$~MeV/c for exciting a 
$\Sigma N$ pair. Although $\Sigma$ hyperon degrees of freedom are generally 
excluded at LO, they may be entered implicitly for $p_{\Lambda N}\ll 2m_{\pi}$ 
through a CS $\Lambda N\leftrightarrow\Sigma N$ transition contact term, 
such as in Eq.~(\ref{eq:DvH3}), which we relate within a partially conserved 
SU(3)$_{\rm f}$ to CS $NN$ and $\Lambda N$ diagonal contact terms. Doing so, 
we find that the free-space value ${\cal A}^{(0)}_{I=1}\approx -0.015$ 
persists also in the \lamb{4}{H}-\lamb{4}{He} mirror hypernuclei. 

\noindent 
{\bf Model.}~
The LO \nopieft~interaction for nucleons and $\Lambda$ hyperons consists 
of two-baryon $BB$ and three-baryon $BBB$ $s$-wave contact terms shown 
schematically in Fig.~\ref{fig:LECs}, (a) and (b). These contact terms are 
given by CS potentials of the form 
\begin{equation} \label{eq:V2} 
  V^S_{B_1B_2} = C^{S}_{B_1B_2}(\lambda)\,{\cal P}_{S}
  \delta_\lambda(\rvec_{12}), 
\end{equation} 
and 
\begin{equation} 
  V^{IS}_{B_1B_2B_3} = D^{IS}_{B_1B_2B_3}(\lambda)\,{\cal Q}_{IS}
  \sum_{\rm cyc}\delta_\lambda(\rvec_{12})\delta_\lambda(\rvec_{23}). 
\label{eq:V3} 
\end{equation}
Here, the $\lambda$ (fm$^{-1}$) dependence attached to the low energy 
constants (LECs) $C_{B_1B_2}^{S}$ and $D_{B_1B_2B_3}^{IS}$ stands for 
momentum cutoff values, introduced in a Gaussian form to regularize the
zero-range contact terms: 
\begin{equation}
\delta_\lambda(\rvec)=\left(\frac{\lambda}{2\sqrt{\pi}}\right)^3\,
\exp \left(-{\frac{\lambda^2}{4}}\rvec^{\,2}\right),   
\label{eq:gaussian} 
\end{equation}
thereby smearing a zero-range (in the limit $\lambda\to\infty$) Dirac 
$\delta^{(3)}(\rvec)$ contact term over distances~$\sim\lambda^{-1}$. The 
cutoff parameter $\lambda$ may be viewed as a scale parameter with respect to 
typical values of momenta $Q$. To make observables cutoff independent, LECs 
must be properly renormalized. Truncating {\nopieft} at LO and using values 
of $\lambda$ higher than the breakup scale 2$m_{\pi}$, observables acquire 
a residual dependence $O(Q/\lambda)$ which diminishes with increasing 
$\lambda$. Finally, following the coupled-channel study of Ref.~\cite{CGK04}, 
we estimate that excluding explicit $\Sigma$ hyperon degrees of freedom incurs 
a few percent error. 

In Eq.~(\ref{eq:V2}), ${\cal P}_{S}$ projects on $s$-wave $B_1B_2$ pairs 
($NN$ or $\Lambda N$) with spin $S$ associated, for a given cutoff $\lambda$, 
with four two-body LECs $C^{S}_{B_1B_2}$ fitted to low-energy two-body 
observables, e.g., to the corresponding CS $NN$ and $\Lambda N$ scattering 
lengths. Similarly, in Eq.~(\ref{eq:V3}), ${\cal Q}_{IS}$ project on $NNN$ 
or $\Lambda NN$ $s$-wave triplets with isospin $I$ and spin $S$ associated 
with four three-body LECs $D^{IS}_{B_1B_2B_3}$ fitted to given CS averages 
of binding energies: $A$=3 without $\Lambda$, $A$=3,4 with $\Lambda$, 
thereby making the procedure applied in Ref.~\cite{cbg18} explicitly CS. 
To calculate these binding energies, the $A$-body Schr\"{o}dinger equation 
is solved variationally by expanding the wavefunction $\Psi$ in a correlated 
Gaussian basis within a stochastic variational method~\cite{SVa98}. 
Convergence to a level of below 1~keV was verified by increasing the number 
of basis states. The predictive power of CS \nopieft(LO) in the $s$-shell 
was already tested in Ref.~\cite{cbg18} by calculating binding energies of 
$^4$He and \lamb{5}{He} and may soon be tested by comparing the calculated 
binding energy of \lamblamb{5}{H}-\lamblamb{5}{He} \cite{csbgm19}, constrained 
by the \lamblamb{6}{He} binding energy datum, with a forthcoming measurement 
at J-PARC. A \nopieft(LO) approach has been used recently in discussions 
of the \lamb{3}{H} (hypertriton) lifetime \cite{HH19} and of a likely 
$\Lambda nn$ continuum state \cite{SBBGM22}. 

Introducing CSB, the two-body $\Lambda N$ contact terms in $V_{\Lambda N}$, 
Eq.~(\ref{eq:V2}), are modified by specifying nucleons as protons or neutrons: 
\begin{equation}   \label{eq:V2a} 
  C^{S}_{\Lambda N} {\cal P}_{S} \rightarrow 
  (C^{S}_{\Lambda p}\frac{1+\tau_{Nz}}{2}+
   C^{S}_{\Lambda n}\frac{1-\tau_{Nz}}{2}){\cal P}_{S}, 
\end{equation} 
This suggests to define CS and CSB LECs $C^{S}_{\Lambda N}$ and $\delta 
C^{S}_{\Lambda N}$, respectively, as 
\begin{equation}   \label{eq:V2b} 
  C^{S}_{\Lambda N} = \half(C^{S}_{\Lambda p}+C^{S}_{\Lambda n}),
  \quad
  \delta C^{S}_{\Lambda N} = \half(C^{S}_{\Lambda p}-C^{S}_{\Lambda n}). 
\end{equation} 
The $\Lambda N$ interaction assumes then the form 
\begin{equation} \label{eq:V2c} 
 V_{\Lambda N} = \sum_S{(C^{S}_{\Lambda N}+\delta C^{S}_{\Lambda N}\,\tau_{z})
 {\cal P}_S \delta_\lambda(\rvec_{\Lambda N})}. 
\end{equation} 
The CSB part of this potential, given in terms of two-body LECs 
$\delta C^{S}_{\Lambda N}$, is then treated perturbatively with respect to 
the LO CS wavefunction. No three-body CSB LECs are necessary at this order. 
The systematic accuracy of the present CS model calculations is about 
$(Q/2m_{\pi})^2\approx 6$\%, where $Q\approx\sqrt{2M_{\Lambda}B_{\Lambda}}
\approx 66$~MeV/c is a typical momentum scale in \lamb{4}{H}-\lamb{4}{He}. 
The suppression of OPE with respect to the dominant contact-terms contribution 
found in $\chi$EFT CSB calculations, see e.g. Ref.~\cite{HMN21}, leads to 
a similar error estimate $\sim$6\% also for the present CSB calculations. 
Nevertheless, if judged by our residual cutoff dependence, a slightly larger 
error estimate is suggested. The precise determination of LO error requires 
consideration of higher order terms.

\begin{figure}[!t] 
\begin{center} 
\includegraphics[width=0.48\textwidth]{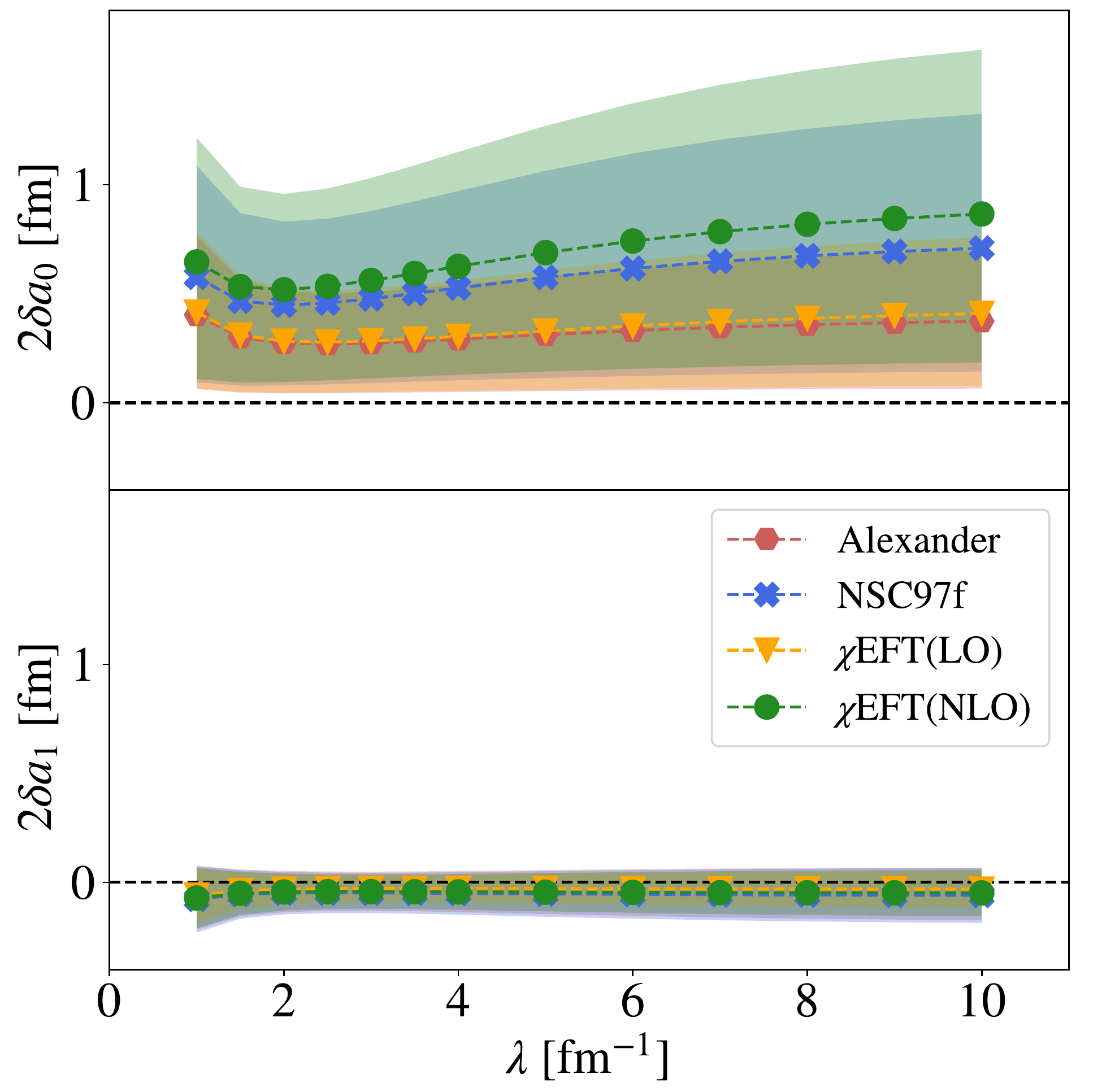} 
\caption{Scattering length differences $2\delta a_S=a_S(\Lambda p) - 
a_S(\Lambda n)$ for $S$=0 (upper) and $S$=1 (lower) as a function 
of cutoff momentum $\lambda$, derived from $\delta C^{S}_{\Lambda N}$ 
LECs extracted here in $A$=4 CSB calculations using $\Lambda N$ 
potential models \cite{Alex68,NSC97,Polinder06,Haiden13}, see text. 
The colored bands represent uncertainties induced by experimental errors in 
$\Delta B_{\Lambda}(0_{\rm g.s.}^+)$ and $\Delta B_{\Lambda}(1_{\rm exc.}^+)$ 
values.} 
\label{fig:delta a} 
\end{center} 
\end{figure}

\noindent 
{\bf Results and discussion.}~
The $\Lambda N$ CSB LECs $\delta C^{S}_{\Lambda N}$, $S$=0,1, were fitted 
pertubatively to the two $A$=4 binding-energy differences $\Delta B_{\Lambda}
(0_{\rm g.s.}^+)$ and $\Delta B_{\Lambda}(1_{\rm exc.}^+)$ shown on the right 
of Fig.~\ref{fig:A=4}. Here we solved the resulting two linear equations for 
$\delta C^{S}_{\Lambda N}$, $S$=0,1, using LO \lamb{4}{H}($0^+,1^+$) 
wavefunctions generated by using CS LECs exclusively. 
Using \lamb{4}{He}($0^+,1^+$) wavefunctions instead, leads to essentially the 
same result. Readjusting the CS three-body LECs $D^{IS}_{\Lambda NN}$ within 
given experimental errors of $B_{\Lambda}$s in the $A=3,4$ hypernuclear 
systems incurs only a few percent uncertainties in the fitted CSB LECs. 
The derived CSB LECs $\delta C^{S}_{\Lambda N}$, of order 1\% of the 
respective $\Lambda N$ CS LECs $C^{S}_{\Lambda N}$, were used in 
a distorted-wave Born approximation to produce $\Lambda N$ scattering 
length differences $\delta a_{\Lambda N}=\half(a_{\Lambda p}-a_{\Lambda n})$. 
Since there is no direct experimental extraction of $\Lambda N$ 
scattering lengths, we used several model estimates for $a_S(\Lambda N)$ 
as input to our calculations. These models, including $\chi$EFT(LO) 
\cite{Polinder06} and $\chi$EFT(NLO)~\cite{Haiden13} used in recent $A$=4 
CSB calculations \cite{GG16,HMN21}, respectively, are listed in an inset to 
Fig.~\ref{fig:delta a} and cited in its caption. The figure shows calculated 
values of 2$\delta a_S$ as function of the cutoff momentum $\lambda$ in 
these $\Lambda N$ interaction models. In agreement with Ref.~\cite{HMN21} 
we find that CSB hardly affects the spin triplet $a_{S=1}$, whereas the 
singlet $a_{S=0}$ of order $\sim(-2.5\pm 0.5)$ fm is affected strongly, making 
$|a_0(\Lambda n)|$ larger by about 0.5 fm than $|a_0(\Lambda p)|$, roughly 
in proportion to $|a_0|$. The dominance of $S$=0 CSB is shown below to arise 
naturally from SU(3)$_{\rm f}$ considerations. Interestingly, going to pure 
neutron matter the size of the (attractive) spin averaged $\Lambda N$ 
scattering length $(3a_1+a_0)/4$ increases by only $\sim$10\% from its 
approximately 2~fm value in symmetric nuclear matter, thereby somewhat 
aggravating the `hyperon puzzle' \cite{GanLon17,Bomb17,TF20} in neutron star 
matter. 

To check the present extraction of $a_S(\Lambda p)-a_S(\Lambda n)$ 
we also applied a similar procedure to the nuclear $NN$ case. 
We verified, successfully, that the experimentally derived CSB 
difference of $a_0(pp)-a_0(nn)$=1.6$\pm$0.6~fm in the $(NN)_{S=0}$ 
sector can be obtained in \nopieft(LO) from the $A$=3 nuclear `datum' 
$\Delta B_{\rm CSB}(^3$H-$^3$He)=67$\pm$9~keV \cite{MM01}. Details will 
be given elsewhere. 

\begin{figure*}[!t] 
\centering 
\includegraphics[width=0.98\linewidth]{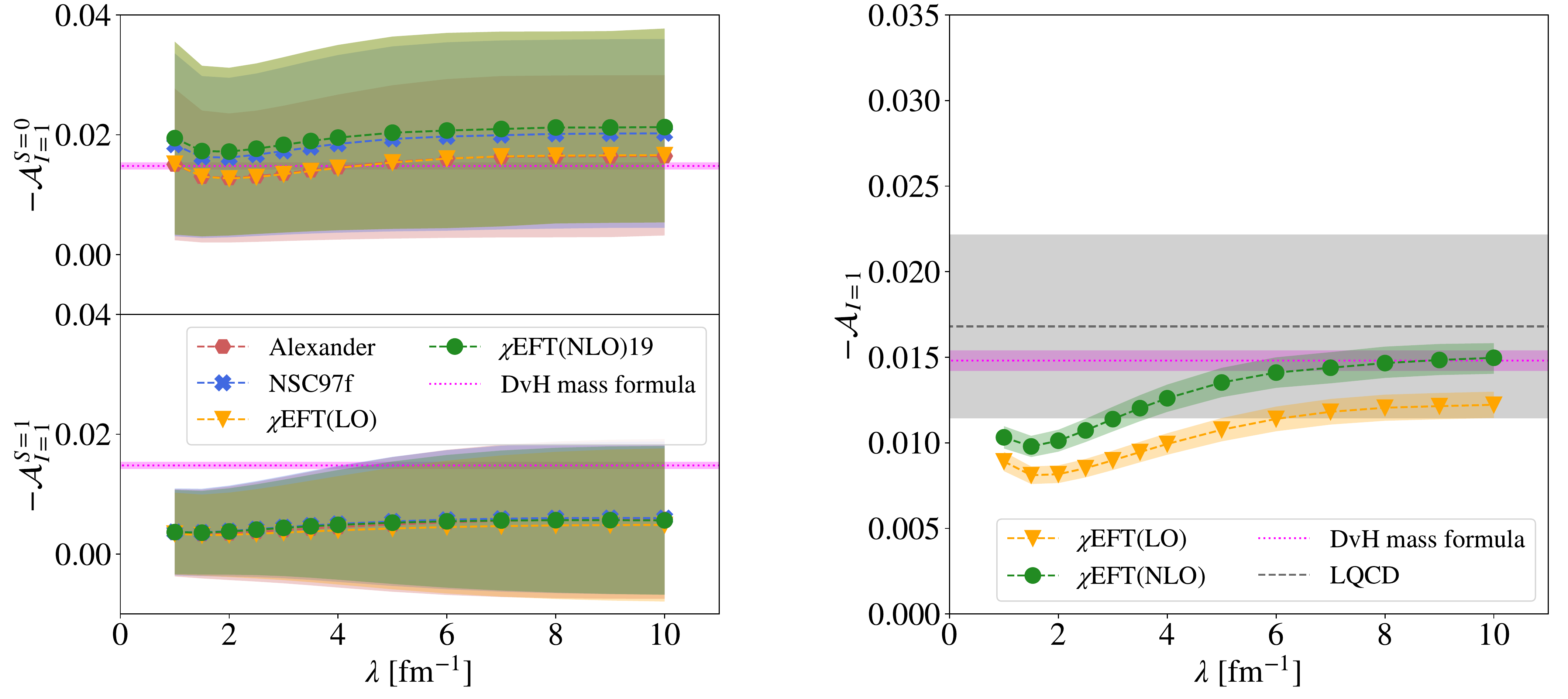} 
\caption{Left: \nopieft\; estimates for the in-medium DvH amplitudes 
$-{\cal A}_{I=1}^S=(\sqrt{3}/2)\delta C^S_{\Lambda N}/C^S_{\Sigma N,\Lambda N}
$ ($S$=0, upper; $S$=1, lower), with $\delta C^S_{\Lambda N}$ derived 
by fitting $\Delta B_{\Lambda}(0_{\rm g.s.}^+)$ and $\Delta B_{\Lambda}
(1_{\rm exc.}^+)$, plotted as a function of the cutoff momentum 
$\lambda$ and specified in an inset by the $\Lambda N$ interaction 
model \cite{Alex68,NSC97,Polinder06,Haiden13} input. 
Colored bands provide uncertainties caused by $\Delta B_{\Lambda}$ 
input values. Right: $-{\cal A}_{I=1}$ values from 
$\Delta E_{\gamma}=E_{\gamma}$(\lamb{4}{He})$-E_{\gamma}$(\lamb{4}{H}), 
see text. Colored bands provide uncertainties caused by that of 
$\Delta E_{\gamma}$. The horizontal colored intervals mark, in pink, 
the DvH \cite{DvH64} value of $-{\cal A}^{(0)}_{I=1}=0.0148\pm 0.0006$ 
from Eq.~(\ref{eq:DvH2}), and in grey the LQCD \cite{LQCD20} value 
$0.0168\pm 0.0054$, see also Table~\ref{tab:DvH}. Note the expanded 
vertical scale on the right with respect to that on the left.} 
\label{fig:DvH} 
\end{figure*} 

Proceeding to the main point of this work, the ansatz Eq.~(\ref{eq:DvH3}), 
we identify $\langle\Lambda N|V_{\rm CSB}|\Lambda N\rangle$ for a given spin 
value $S$=0,1 with the CSB LEC $\delta C^{S}_{\Lambda N}$ extracted directly 
from the \lamb{4}{H}-\lamb{4}{He} spectrum. Similarly, working within the 
framework of \nopieft\; we identify the spin dependent matrix-element 
$\langle\Sigma N|V_{\rm CS}|\Lambda N \rangle$ with a new $\Sigma N
\leftrightarrow\Lambda N$ LEC $C^S_{\Sigma N,\Lambda N}$. 
Following Dover and Feshbach~\cite{DF90}, we use SU(3)$_{\rm f}$ to relate 
$C^S_{\Sigma N,\Lambda N}$ to the $NN$ and $\Lambda N$ CS LECs 
established in the present application of \nopieft(LO): 
\begin{align} 
  C^0_{\Sigma N,\Lambda N} &= -3(C^0_{NN}-C^0_{\Lambda N}),
  \cr
  C^1_{\Sigma N,\Lambda N} &= \;(C^1_{NN}-C^1_{\Lambda N}). 
\label{eq:SU(3)} 
\end{align} 
In the next step, inspired by Eq.~(\ref{eq:DvH3}), we replace the two CSB 
LECs $\delta C^S_{\Lambda N}$ derived from the two binding energy differences 
$\Delta B_{\Lambda}(0^+_{\rm g.s.})$ and $\Delta B_{\Lambda}(1^+_{\rm exc.})$ 
by two in-medium amplitudes ${\cal A}_{I=1}^S$, $S=0,1$, 
defined by 
\begin{equation} 
-{\cal A}_{I=1}^S=(\sqrt{3}/2)\,\delta C^S_{\Lambda N}/C^S_{\Sigma N,
\Lambda N},
\label{eq:DvH_S} 
\end{equation}  
where the CS LECs $C^S_{\Sigma N,\Lambda N}$ are expressed through 
Eq.~(\ref{eq:SU(3)}), thereby eliminating any explicit reference to $\Sigma$ 
hyperon degrees of freedom. The amplitudes ${\cal A}_{I=1}^S$, $S=0,1$, are 
shown on the l.h.s. of Fig.~\ref{fig:DvH} as function of the cutoff $\lambda$, 
for both $S$=0 (upper) and $S$=1 (lower) 2-body spin states, using the same 
$\Lambda N$ interaction models cited in the inset of Fig.~\ref{fig:delta a}. 
These amplitudes exhibit a rather weak dependence on $\lambda$, 
with a common value consistent with (and for $S$=0 close to) the DvH value 
$-{\cal A}^{(0)}_{I=1}=0.0148$ from Eq.~(\ref{eq:DvH2}). 

The results exhibited for ${\cal A}_{I=1}^S$, $S=0,1$, in the left panel 
of Fig.~\ref{fig:DvH} were derived using the relatively imprecise binding 
energy differences $\Delta B_{\Lambda}(0^+_{\rm g.s.})$=233$\pm$92~keV and 
$\Delta B_{\Lambda}(1^+_{\rm exc.})$=$-$83$\pm$94~keV, see Fig.~\ref{fig:A=4}. 
Using instead the considerably more precise single value $\Delta E_{\gamma}=
316\pm 20$~keV obtained from the difference between the two $\gamma$ ray 
energies marked in the figure is, however, insufficient to determine both 
${\cal A}_{I=1}^{S=0}$ and ${\cal A}_{I=1}^{S=1}$. Assuming spin independence 
to start with, a precise value ${\cal A}_{I=1}$ is derived from $\Delta 
E_{\gamma}$ as shown on the r.h.s. of Fig.~\ref{fig:DvH}. For all four 
$\Lambda N$ potential models considered on the l.h.s. the derived value 
(shown for the $\chi$EFT models) is within the LQCD horizontal band for 
cutoff $\lambda \gtrsim 6$ fm$^{-1}$, while for NSC97f \cite{NSC97} and 
$\chi$EFT(NLO) \cite{Haiden13} (and also its 2019 version \cite{HMN20}) 
the derived values do enter the considerably narrower ($\pm 4\%$) DvH 
SU(3)$_{\rm f}$ horizontal band for $-{\cal A}^{(0)}_{I=1}$. All in-medium 
isospin $I=1$ admixture amplitudes $-{\cal A}_{I=1}$ calculated for 
$\lambda\gg 2m_\pi$ exhibit a $\sim\lambda^{-1}$ dependence, and the 
corresponding values extrapolated to the renormalization scale-invariance 
limit $\lambda\to\infty$ are listed in Table~\ref{tab:DvH}. 

\begin{table}[htb] 
\caption{Estimates of the $I$=1 admixture amplitude $-{\cal A}_{I=1}$ 
in the $\Lambda$ hyperon from (i) baryon-number $B$=1 free-space studies of 
DvH and LQCD \cite{DvH64,LQCD20} and (ii) the present \nopieft(LO) $B$=4 
\lamb{4}{H}-\lamb{4}{He} CSB study, using Eq.~(\ref{eq:DvH3}) and input from 
$\Lambda N$ $\chi$EFT models. The $B$=4 values are extrapolations to the 
renormalization scale invariance limit $\lambda\to\infty$ and their listed 
uncertainties reflect primarily input data uncertainties.} 
\begin{center}
\label{tab:DvH}
\begin{tabular}{l c c c }
\hline \hline
Method/Input & $B$ & & $-{\cal A}_{I=1}$ \\ 
\hline
SU(3)$_{\rm f}$ \cite{DvH64} & 1 & & $0.0148\pm 0.0006$ \\
LQCD \cite{LQCD20} & 1 & & $0.0168\pm 0.0054$ \\
\nopieft(LO)/$\chi$EFT(LO) \cite{Polinder06} & 4 & & $0.0139\pm 0.0013$ \\
\nopieft(LO)/$\chi$EFT(NLO) \cite{Haiden13}  & 4 & & $0.0168\pm 0.0014$ \\ 
\hline \hline 
\end{tabular} 
\end{center} 
\end{table} 

In view of its success, the above procedure can also be applied in reverse. 
That is, substituting the DvH value ${\cal A}^{(0)}_{I=1}$ from 
Eq.~(\ref{eq:DvH2}) and the SU(3)$_{\rm f}$ relations Eq.~(\ref{eq:SU(3)}) in 
Eq.~(\ref{eq:DvH_S}), one extracts the CSB LECs $\delta C^S_{\Lambda N}$ and 
evaluates the $A=4$ CSB splittings $\Delta B_{\Lambda}(0_{\rm g.s.}^+)$ and 
$\Delta B_{\Lambda}(1_{\rm exc.}^+)$. Doing so, we find that the $\Delta 
B_{\Lambda}$ values marked in Fig.~\ref{fig:A=4} are reproduced within their 
experimental error. We also find that (i) the resulting LECs $\delta C^0_{
\Lambda N}$, $\delta C^1_{\Lambda N}$ have opposite signs, and (ii) $|\delta 
C^0_{\Lambda N}| \gg |\delta C^1_{\Lambda N}|$, implying that CSB acts 
predominantly on the spin $S$=0 channel. 

Our results suggest strongly that the $\Lambda$'s $I=1$ isospin impurity 
of magnitude $\approx$1.5\% in free space is upheld in the $A=4$ $\Lambda$ 
mirror hypernuclei. This conclusion appears natural in the context of partial 
restoration of chiral symmetry in dense matter, since there is no direct link 
known to us between $\Lambda -\Sigma^0$ mixing and chiral symmetry breaking; 
for example, the non-strange quark-mass difference $m_d-m_u$ does not enter 
the SU(3)$_{\rm f}$ mass mixing matrix element $M_{\Sigma^0\Lambda}$, 
Eq.~(\ref{eq:DvH1}). 

\noindent 
{\bf Closing remarks.}~
{\nopieft}(LO) applications to few-body hypernuclei are normally limited to 
$N$ and $\Lambda$ degrees of freedom. To consider hypernuclear CSB, the set 
of two-body and three-body $s$-wave CS LECs shown schematically in (a) and 
(b) of Fig.~\ref{fig:LECs} was extended, adding two $S=0,1$ $\Lambda N$ CSB 
LECs which were then fitted to the two experimentally available CSB $A=4$ 
$\Delta B_{\Lambda}$ values. The resulting CS broken values of the $\Lambda N$ 
scattering lengths shown in Fig.~\ref{fig:delta a} come out then practically 
the same as derived within a more involved $\chi$EFT(NLO) approach 
\cite{HMN21} that includes additional $\Sigma$ hyperon and pseudoscalar octet 
meson degrees of freedom. 

Apart from suggesting an economical way to evaluate CSB in strange matter, 
we were able to show that CSB is linked uniquely to the $I=1$ isospin impurity 
${\cal A}^{(0)}_{I=1}$ of the dominantly $I=0$ $\Lambda$ hyperon, provided (i) 
the ansatz (\ref{eq:DvH3}) is adopted and (ii) the $\langle\Sigma N|V_{\rm CS}
|\Lambda N\rangle$ transition matrix element in (\ref{eq:DvH3}) is related by 
SU(3)$_{\rm f}$ to the $\langle NN|V_{\rm CS}|NN\rangle$ and $\langle\Lambda 
N|V_{\rm CS}|\Lambda N\rangle$ matrix elements, each near its own threshold. 
Having managed to avoid introducing explicitly $\Sigma$ hyperon degrees of 
freedom into the employed \nopieft(LO) scheme, we are spared of trying to 
impose SU(3)$_{\rm f}$ symmetry simultaneously on all $\Lambda N$, $\Sigma N$ 
and $\Lambda N\leftrightarrow\Sigma N$ LECs fitted to low-energy scattering 
and reaction data, proven impossible at LO in $\chi$EFT \cite{Polinder06} 
as discussed recently~\cite{HMN20} upon introducing a new $\chi$EFT(NLO)19 
version. It is worth mentioning that two $\Sigma N$ channels, 
($I$=$\frac{3}{2}$ $J^{\pi}$=$1^+$) and ($I$=$\frac{1}{2}$ $J^{\pi}$=$0^+$), 
are plagued by `Pauli forbidden' six-quark configurations, say $uuuuds$ for 
$I_z$=$\frac{3}{2}$ of the former channel \cite{OSY87}, providing thereby 
a specific mechanism beyond the scope of SU(3)$_{\rm f}$. 

The usefulness of the CSB approach outlined here for the $A=4$ mirror 
hypernuclei, where precise $\gamma$ ray data exist, should be tested 
in heavier hypernuclei when similarly precise CSB data become available. 
However, phenomenological arguments regarding $\Sigma$ hyperon admixtures 
in $\Lambda$ hypernuclei \cite{gal15,GM13} lead us to believe that CSB 
splittings of $1s_{\Lambda}$ mirror levels decrease quickly with $A$, 
making such tests more difficult but nonetheless challenging.  

\noindent 
{\bf Acknowledgment}~
AG would like to thank Johann Haidenbauer and Wolfram Weise for instructive 
discussions on EFT approaches to strange few-body systems. The work of MS and 
NB was supported by the Pazy Foundation and by the Israel Science Foundation 
grant 1086/21. Furthermore, the work of AG and NB is part of a project funded 
by the European Union's Horizon 2020 research and innovation programme under 
grant agreement No. 824093.

\end{document}